# Topological Catenation-induced Pore Size in 2D Olympic Network

*Wenbo Zhao[1,2], Guojie Zhang\*[,3], Hong Liu\*[,1,2]*

[1] Key Laboratory of Theoretical Chemistry of Environment Ministry of Education, School of Environment, South China Normal University, Guangzhou 510006, China;

[2] Guangdong Provincial Key Laboratory of Chemical Pollution and Environmental Safety, School of Environment, South China Normal University, Guangzhou 510006, China;

[3] School of Chemistry and Chemical Engineering, Guangzhou University, Guangzhou 510006, China.

**ABSTRACT:** Assemblies of ring DNA and proteins—such as kinetoplast DNA (kDNA) and the chainmail-like network of the HK97 bacteriophage capsid—form "Olympic" topologies that govern the dynamics of gene-regulatory proteins, cellular metabolites, and viral genomes. Yet the pore sizes that control transport and diffusion in these topologically linked networks remain poorly quantified. Here, using coarse-grained simulations of idealized square (SQR) and hexagonal (HEX) lattice-catenated two-dimensional Olympic networks, we measure pore size as the radius of the largest rigid tracer that can pass through an aperture and delineate how backbone bending stiffness and topological tension regulate this structure. We identify two competing regulators: (i) ring entropic elasticity, which enlarges pores at low stiffness, and (ii) ring rotation angle, which reduces pore size at high stiffness. Their competition yields bimodal pore-size distributions over a defined parameter range. Network geometry further modulates spatial correlations in pore size and the pore-size autocorrelation time, predicting distinct propagation of perturbations in isostatic (SQR) versus hypostatic (HEX) architectures. These results provide testable predictions for enzyme diffusion kinetics and cargo release in natural or synthetic catenane DNA networks and mechanically interlocked capsid proteins, and offer a theoretical framework for interpreting FRAP (fluorescence recovery after photobleaching) and SMT (single-molecule tracking) measurements in topologically catenated biopolymer networks.

# 1. INTRODUCTION

Topology has become one of the emerging key concepts in natural science during the past decades. In the field of soft condensed matter, topology and topological effect (i.e., topological constraints) play a pivotal role in determining statics, dynamics and properties/functions of polymers in both man-made soft materials (e.g., elastomers, hydrogels, etc.) and life systems (e.g., DNA molecules in cell nucleus, proteins, etc.)[1–3]. For instance, topological constraints in dense linear polymers result in emergence of polymer entanglements, which are responsible for appearance of distinct dynamic modes (i.e., one-dimensional reptation) of polymer chains and thus nontrivial and promising viscoelastic properties of these materials. Ring or cyclic polymer is a special class of macromolecules characterized by absence of chain ends along the chain backbone, and thus represents a kind of topologically complex but very interesting systems. Similarly, circular topology and corresponding topological constraints in ring polymers has a profound effect on conformational statistics and dynamics of polymer chains with unique features being absent in its linear counterpart. Understanding the relationship between molecular topology and structural and physical properties of topologically more complex polymer systems remains one of promising research frontiers in soft matter physics[4–9].

Interlocked ring polymers, which are formed by concatenation of rings, have been attracting considerable attention not only due to their intrinsic beauty in molecular topology but also potential applications in many aspects. More interestingly, mechanically interlocked rings, e.g., catenated DNA molecules and proteins, are rather ubiquitous in life systems[9–12]. Regarding protein catenanes, interlocked ring architectures are found in viral systems such as HK97 capsids, where lattice connectivity and prestress provide a two-dimensional capsid-like context for ring-defined pores[13,14]. The interlocked DNA rings are often encountered during DNA replication and recombination, and when ring-shaped protein complexes (e.g., SMC) restructure chromosomes, i.e., when topoisomerases both generate and resolve these interlocks, dynamically regulating DNA topology in cells[15–19]. Catenation of circular DNAs is also observed for mitochondrial/plasmid DNA and in recombination intermediates in bacteria and eukaryotes. kinetoplast DNA (kDNA) in trypanosome mitochondria could be one of the most fascinating examples of naturally occurring interlocked ring polymers. kDNA can be seen a two-dimensional "Olympic" network composed of thousands of short DNA minicircles (~0.5 – 2.5 kb) mechanically interlocked into a chain-mail–like sheet that supports mitochondrial genome maintenance and RNA editing[20–22]. Its function engages multiple enzymes, notably topoisomerase II (TOP II), which mediates decatenation or recatenation during replication and network remodeling. More broadly, catenated topologies have been speculated to modulate meiotic processing and viral DNA integration, with implications for genome stability and gene-editing outcomes[23,24].

Nevertheless, the biological advantage of interlocked topology in kDNA molecules still remains unclear. Usually, the functional performance of biological processes in polymer networks is ultimately governed by the diffusion dynamics of guest molecules that navigate the network pores[25–28]. For instance, intracellular events — DNA replication, transcription, and protein biosynthesis — are all diffusion-limited enzymatic reactions whose efficiencies are intrinsically linked to the underlying pore architecture: solute mobility is constrained by both mesh size and the network's topological restrictions[29–31]. It has been speculated that the interlocking structure of kDNA provides genomic stability, e.g., mechanical stability the organelle, which could be reminiscent of similar mechanical stability postulated enhanced by catenation network in HK97 capsids. Therefore, understanding effect of the catenation topology upon conformational and physical properties, such as pore size within two-dimensional catenation network, is fundamentally desirable. Due to current limitations on experimental characterization techniques, unfortunately, there has been little progress made along this line.

Here we develop such a framework using coarse-grained molecular-dynamics simulations of two idealized two-dimensional "Olympic" lattices. We model a two-dimensional catenane member with periodic boundary conditions to probe bulk behavior while avoiding edge and finite-size effects. Given that kDNA and other interlocked topological molecular structures in biological systems predominantly exist in bounded forms, we simulated systems with edge effects as a comparative case. Two geometries are considered—a near-isostatic square (SQR) lattice and a hypostatic hexagonal (HEX) lattice—so that Maxwell constraint counting can be used to interpret geometry-dependent responses. Pore size is defined operationally as the largest rigid tracer FRAP, SMT or single-particle tracking. We scan two control knobs: backbone bending stiffness $K_{\text{bend}}$ and a dimensionless topological tension λ.

Our calculations reveal that two governing mechanisms compete to set pore size: (i) entropic elasticity of flexible rings, which expands pores at low $K_{bend}$, and (ii) rotational occupancy of stiffer rings, which invades the pore cross-section and shrinks the accessible aperture at higher $K_{bend}$. The balance of these effects gives a bimodal pore-size distribution for certain ($K_{bend}$, $\lambda$), while increasing $\lambda$ suppresses rotational freedom and homogenizes pores. Constraint topology further modulates collective behavior: SQR (near-isostatic) networks are highly sensitive to chain-level parameters and display negative correlations between orthogonal pore dimensions with shorter memory (faster disturbance propagation), whereas HEX (hypostatic) networks—with extra floppy modes—respond more slowly and exhibit longer pore self-correlation times. Together, these results provide testable biophysical predictions for the scaling of tracer diffusivity and enzyme accessibility with $K_{bend}$ and $\lambda$ in the interior of catenated assemblies, offering a baseline to interpret FRAP recoveries, size-dependent single-particle diffusivities, and diffusion-limited reaction rates—without making system-specific claims about biological systems.

## 2. MODEL AND SIMULATION DETAILS

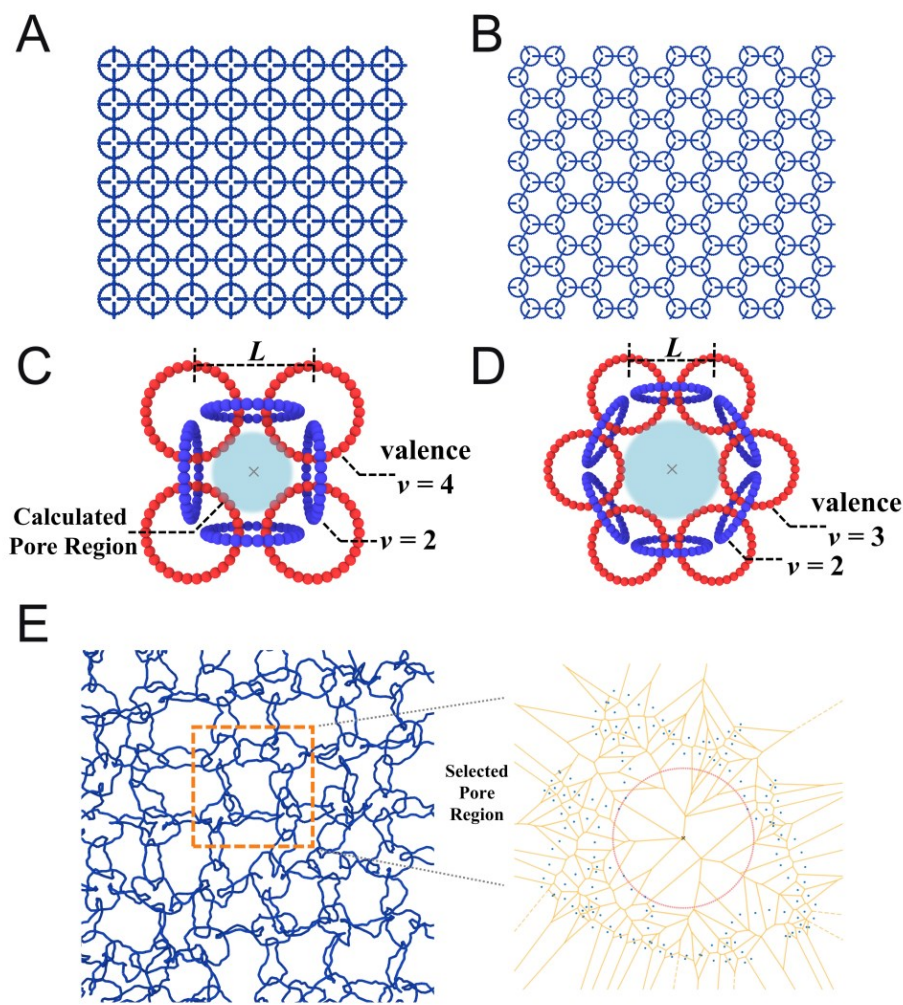

**Figure 1.** **(A)** Snapshot of the SQR network. **(B)** Snapshot of the HEX network. **(C)** A square unit in the SQR network, with red rings ($v = 4$) and blue rings ($v = 2$). The topological tension strength $\lambda$, is defined by the initial centroid distance L, between two adjacent red rings within a unit, with $\lambda = 3.28\ \varepsilon/\sigma$. The light blue region highlights the pore selected for analysis in the SQR network. **(D)** A hexagonal unit in the HEX network, with red rings ($v = 3$) and blue rings ($v = 2$), where $\lambda = 3.28\ \varepsilon/\sigma$. The light blue region highlights the pore selected for analysis in the HEX network. **(E)** Left panel shows the relaxed conformation of the SQR network at $K_{bend} = 0$ $\varepsilon/\sigma^2$ and $\lambda = 3.28 \ \varepsilon/\sigma$. The right panel presents an algorithmic schematic for pore size characterization, where the yellow lines denote Voronoi boundaries, the blue dots represent polymer particles, and the red dashed circle indicates the maximum inscribed circle corresponding to the pore size. The center of the circle is marked by a black cross.

In this study, we constructed two idealized two-dimensional chain-linked network models. As shown in Fig. 1A, the first is a square-lattice network, hereafter referred to as the SQR network, while Fig. 1B illustrates a hexagonal-lattice network, hereafter referred to as the HEX network. The rings within each network are distinguished by their relative spatial orientations. For example, Fig. 1C depicts a square unit in the SQR network, where the red rings lie parallel to the XY-plane and the blue rings are oriented perpendicular to the XY-plane. Similarly, Fig. 1D shows a hexagonal unit of the HEX network, in which the classification of red and blue rings follows the same convention.

In our study, we employed the conventional Kremer-Grest model based coarse-grained molecular dynamics simulation method[32]. We set the ring size as M = 30, i.e., each ring consists of 30 coarse-grained particles with a diameter of $\sigma$.

The non-bonded interactions between polymer monomers, as well as any two contacting particles were described by a truncated Lennard-Jones potential[33] (also called as WCA, i.e., Weeks-Chandler-Andersen potential), given as:

$$U_{LJ}(r) = \begin{cases} 4\varepsilon \left[ \left(\frac{\sigma}{r}\right)^{12} - \left(\frac{\sigma}{r}\right)^{6} \right] + \varepsilon \ ; & \left(r \leq 2^{\frac{1}{6}}\sigma\right) \\ 0 & ; \left(r > 2^{\frac{1}{6}}\sigma\right) \end{cases}. \quad (1)$$

In Equation 1, $\varepsilon$ represents the depth of the potential well, $r$ denotes the distance between two polymer particles, and $\sigma$ serves as the unit of length. To simplify the simulation, we set $\varepsilon = \sigma = m$ (the mass of a particle) = 1.0. The cutoff radius $r_c$ is set to $2^{1/6}\sigma$ in the simulation. Consequently, only repulsive interactions remain between any two polymer particles.

The bonded interactions between adjacent monomers within the ring polymer are characterized using the finitely extensible nonlinear elastic (FENE) potential:

$$U_{FENE} = -\frac{1}{2} k R_0^2 \ln\left(1 - \frac{r^2}{R_0^2}\right). \quad (2)$$

In Equation 2, $k = 30\varepsilon/\sigma^2$, and $R_0 = 1.5\sigma$. Under these parameters, the equilibrium bond length of the system is maintained at $l_0 = 0.97\sigma$ according to the conventional Kremer-Grest model[32]. The above settings apply to the bonded interactions between polymer particles within all ring polymers in the system. The stiffness of the ring polymer chains is described by a standard harmonic bond angle potential in Equation 3 as:

$$U_{angle} = \frac{1}{2} K_{bend} (\theta - \theta_0)^2. \quad (3)$$

Here, $K_{bend}$ is the bending force constant, with units of $\varepsilon/\sigma^2$. $\theta$ represents the bond angle formed by three consecutive monomers along the polymer chain, and $\theta_0$ is the equilibrium bond angle. For the ring polymer system simulated in this study, the equilibrium bond angle is set to 180°.

All molecular dynamics simulations were performed using the GALAMOST simulation package[34]. The simulations utilized the NVT ensemble with Langevin dynamics, based on energy units $\varepsilon$, length units $\sigma$, and mass units $m$. The system temperature was set to $T = \varepsilon/k_B$, where $k_B$ is the Boltzmann constant. The time step was set to $dt = 0.005\sigma(m/\varepsilon)^{1/2}$. Due to the presence of a large number of catenane structures in the system, an extended equilibration simulation of $1 \times 10^8$ time steps was first performed to relax the system and eliminate the effect of the initial configuration. Subsequently, an additional $2 \times 10^8$ time steps were conducted to collect data. The periodic boundary conditions are applied in all three directions. By slightly regulating the simulation box length $L_x$ and $L_y$, we are able to achieve that $L_x$ and $L_y$ are adaptive as integer multiple of the size of initially arranged ring polymers.

As illustrated in Figs. 1C and 1D, three adjustable parameters were employed in this study: the polymer backbone stiffness $K_{bend}$, the valence of ring connections $v$ and the network topological tension $\lambda$. In the simulations, the backbone stiffness was varied by setting $K_{bend}$ = 0, 5, 10, 20, 30, 40, 50 $\varepsilon/\sigma^2$, with the detailed definition provided in Eq. (3). The network topological tension $\lambda$ was defined as the average tension acting on each polymer ring, with values set to $\lambda$=3.18 $\varepsilon/\sigma$, 3.22 $\varepsilon/\sigma$, 3.28 $\varepsilon/\sigma$ and 3.35 $\varepsilon/\sigma$. To eliminate local volumetric effects arising from network curvature, we placed $N_s$ = 3072 rings in the SQR system and $N_H$ = 2940 rings in the HEX system. This configuration ensured that the lateral dimensions of the two-dimensional networks were significantly larger than the chosen value of $L$. Comprehensive molecular dynamics simulations were then performed on both the SQR and HEX networks, covering the full range of backbone stiffness and topological tension values. For comparison, SQR and HEX networks with boundaries were also simulated under different values of backbone stiffness.

The topology of a network profoundly affects its stability and mechanical properties. On a molecular scale, different polymer monomers form various topological structures through processes such as cross-linking, entanglement and cyclization, which in turn influence the macroscopic mechanical properties of polymer materials. At a larger scale, the different connection methods of cables and beams in buildings also determine structural stability. Tracing back to 1864, James Clerk Maxwell introduced the concept of networks made up of rigid rod frameworks and connecting nodes, establishing a link between the number of frames and nodes with the network's stability[35]. Moreover, networks built from rigid rods and connecting points can be directly extended to networks composed of mass points and bonds. Specifically, the Maxwell count can be calculated using the following formula to assess the stability of a network:

$$z = dN_s - N_b. \tag{4}$$

In Equation 4, $z$ denotes the Maxwell count; $d$ represents the spatial dimension of the network; $N_s$ is the number of mass points in the network; and $N_b$ is the number of bonds. However, the Maxwell count described by Equation 4 does not apply in all cases, such as for networks under periodic boundary conditions. For solving this problem, researchers have proposed a more general Maxwell relation:

$$z = dN_s - N_b = N_0 - N_{ss}. \tag{5}$$

In Equation 5, $N_0$ represents the number of zero modes in the system, which can also be considered the system's degrees of freedom. $N_{ss}$ stands for the number of States of Self-Stress in the system, also known as the number of constraints. Zero modes refer to the ability of mass points to move without compressing or stretching the bonds, and such movements do not generate any elastic potential energy, revealing the flexibility inherent in the network topology. On the other hand, States of Self-Stress describe the distribution of internal stress: although bonds may be under compression or tension, the forces between them balance out at the connection points, producing no net external force, which indicates the system's rigidity.

By calculating the Maxwell relation of the network using Equation 5, we obtain three distinct cases:

$$z \begin{cases} < 0 & \text{hyperstatic} \\ = 0 & \text{isostatic} \\ > 0 & \text{hypostatic} \end{cases}. \tag{6}$$

In Equation 6, when $z < 0$, the network is considered hyperstatic, meaning that its degrees of freedom are fewer than its constraints, resulting in high rigidity. When $z > 0$, the network is hypostatic, as it has more degrees of freedom than constraints and is rich in zero modes, which makes it flexible. When $z = 0$, the degrees of freedom and constraints are exactly balanced, placing the network at the critical point between rigidity and flexibility; we call such a network isostatic. In systems with non-periodic boundaries, isostatic networks can exhibit a pronounced soft–hard contrast at their edges, meaning that the mechanical properties at the boundaries differ significantly, and this state remains robust against material partitioning and internal defects. Isostatic networks are widespread in nature—from the packing of sand and nanoparticles to network structures in silicate glasses and polymer crystallites—endowing these materials with many unique physicochemical properties[36–38]. With continued research on soft matter systems and the Maxwell relation, Kane and Lubensky integrated the concept of topological states from topological insulators and superconductors with the Maxwell mechanical framework, thereby proposing a new theory of topological mechanical metamaterials[39,40]. The work of Zhou and colleagues confirmed that this theory also applies to soft matter systems; their results indicate that disordered fiber networks can exhibit highly non-uniform mechanical properties, which play a unique role in the cytoskeleton and the extracellular matrix[41,42].

Turning our attention back to the SQR and HEX networks shown in Fig. 1A and Fig. 1B, if we treat the $v = 2$ elements in both networks as mass points and the $v = 3$ and $v = 4$ rings as spring bonds connecting these points, we obtain a two-dimensional network formed by a square lattice and another network composed of hexagonal lattices arranged in a honeycomb pattern. Under periodic boundary conditions, calculating the Maxwell count using Equation 5 requires selecting a unit cell of the network as the smallest element and determining the number of mass points and bonds within that cell. For the two-dimensional network shown in Fig. 1A, each mass point connects to four nearest neighbors via bonds, and since each bond links two mass points, each unit cell on average contains 1 mass point and 2 bonds. Substituting these values into the formula yields $z = dN_s - N_b = 2 \times 1 - 2 = 0$, classifying this network as isostatic. Similarly, for the two-dimensional network shown in Fig. 1B, each mass point connects to three nearest neighbors, and each bond still connects two mass points. On average, each unit cell contains 2 mass points and 3 bonds, leading to $z = dN_s - N_b = 2 \times 2 - 3 = 1$, which indicates a hypostatic network. Correlating the geometric mapping discussed earlier, the SQR network can be regarded as isostatic and the HEX network as hypostatic.

In the Olympic network, the unique topology of connections that do not involve chemical bonds makes the standard algorithm unsuitable because it relies on calculating the shortest path along chemical bonds. Consequently, it is still a challenging task on accurately calculating the pore size in a specified region of the Olympic network. In this study, we developed a method to calculate the precise dimensions of pores in specific regions of the network. We define the pore size as the radius of the largest spherical rigid particle that can pass through the pore, a definition inspired by Torquato et al. on void size. This definition allows the calculated pore size to be used for further investigations into physical properties such as the diffusion characteristics of the network. Our method utilizes the Graham scan algorithm to identify and index the particles that form the boundary of a specific pore region. Combined with the Voronoi diagram algorithm, this approach is able to quickly identify the center of the largest spherical particle defined above, thereby determining its precise radius. In Fig. 1E, we selected a single quadrilateral unit within a fully relaxed and equilibrated SQR topological network as a representative pore and present a schematic of the algorithm. Particles forming the network are marked with blue dots, showing their distribution in the XY-plane. The Voronoi boundaries of each particle are outlined in yellow lines, partitioning the Voronoi space for each particle, which represents the nearest spatial region to that particle. The intersections of all boundaries, corresponding to the centers of the largest spherical particles as defined above, are marked with black crosses. By calculating the coordinates of these intersections and measuring the distances to all surrounding particles, we identify the largest empty circle and its radius. This circle, corresponding to the sphere's Great Circle, is depicted in Fig. 1E with a red dashed line. Through simulations, we demonstrate that this method is robust and can be extended to other network systems with complex topological structures.

## 3. RESULTS AND DISCUSSION

### pore size

We begin our discussion with the pore size. Referring to the studies by Meng and Prakasam *et al.* on the mechanical properties and pore structures of Olympic networks, it is recognized that the tunable porosity in such materials primarily arises from the "large pores" enclosed by polymer rings interconnected via linker chains, which form the network framework. In contrast, the intrinsic voids within individual polymer rings contribute minimally to the functional pore structure of the material. Based on this understanding, the present study focuses exclusively on the large pores formed by ring assemblies within each network lattice unit, as illustrated by the light blue regions in Fig. 1C and 1D, while excluding those formed solely by the internal ring geometry. To visually illustrate the influence of polymer chain stiffness ($K_{bend}$) and topological tension strength ($\lambda$) on the pore size, we analyzed the coordinates of all particles forming the pores from the dynamic trajectories generated in the simulations. Based on the centroids of the pore-forming particles, we normalized all coordinates and constructed a 2D grid of 100×100 cells in the XY directions. For each grid cell, we calculated the particle density and visualized the results using color gradient heatmaps, as shown in Fig. 2. Fig. 2A presents the particle density heatmaps for the SQR system under three topological tension strength conditions $\lambda$ = 3.35 $\varepsilon/\sigma$ and three polymer stiffness values ($K_{bend}$ = 0, 5, 50 $\varepsilon/\sigma^2$), resulting in three heatmaps on top. Similarly, and the next line displays the corresponding particle density heatmaps for the HEX system under the same parameter space. The heat map illustrating variations with topological tension will be presented in the Supplementary Information. In both, the chain stiffness increases from left to right, while the topological tension strength increases from bottom to top. The color bar indicates particle density, transitioning from lower values in blue to higher values in red. By comparing the heatmaps in Fig. 2A, it can be observed that the SQR and HEX systems exhibit similar evolutionary trends with varying $K_{bend}$ and $\lambda$. When the topological tension strength remains constant, an increase in chain stiffness leads to a significant reduction in the blue regions of the pore area, indicating a decrease in pore size. Conversely, when the chain stiffness is kept constant, an increase in $\lambda$ results in a noticeable expansion of the blue regions, demonstrating that a higher topological tension strength causes the pore size to increase.

Exploring the variation in pore size through changes in the blue regions of the heatmaps provides a visually intuitive understanding of the influences by ring bending rigidity or topological tension. However, more detailed quantitative analysis regarding pore size is still necessary for further in-depth understanding the formation of these pores.

Based on the pore-size calculation algorithm described above, we evaluated the pore size r of both the SQR and HEX network structures under different polymer backbone stiffness values and topological tension strengths. We also calculated the pore sizes of the two networks in the absence of topological tension. Fig 2C and 2D present the variation of r as a function of backbone stiffness and topological tension for the SQR and HEX networks, respectively. From the similar trends observed in Fig. 2C and 2D, it is evident that, regardless of the presence or absence of tension in the network, the influence of backbone stiffness on the r is qualitatively similar for both networks. Specifically, under constant applied tension, the colored data points in the figures show that the r decreases with increasing backbone stiffness, and this reduction tends to saturate when the stiffness reaches $K_{bend}$ = 30 $\varepsilon/\sigma^2$. We attribute this behavior to the fact that, at lower stiffness, polymer rings adopt coiled conformations that occupy less volume within the pore, resulting in larger pore sizes, whereas at higher stiffness, the rings tend toward circular conformations that occupy more pore space, thereby reducing the effective pore size. Furthermore, comparing the pore sizes of the two networks at the same backbone stiffness reveals that the pore size increases with increasing tension. This effect likely arises because polymer rings are stretched into more elongated shapes under larger applied tension, occupying less pore volume and thereby leading to an increase in pore size. In contrast, as shown in Fig. 2E, in the absence of tension, the pore size increases with increasing backbone stiffness in both the SQR and HEX networks, opposite to the trend observed under tension. This latter trend is consistent with previously reported behavior of pore sizes in cross-linked polymer networks.

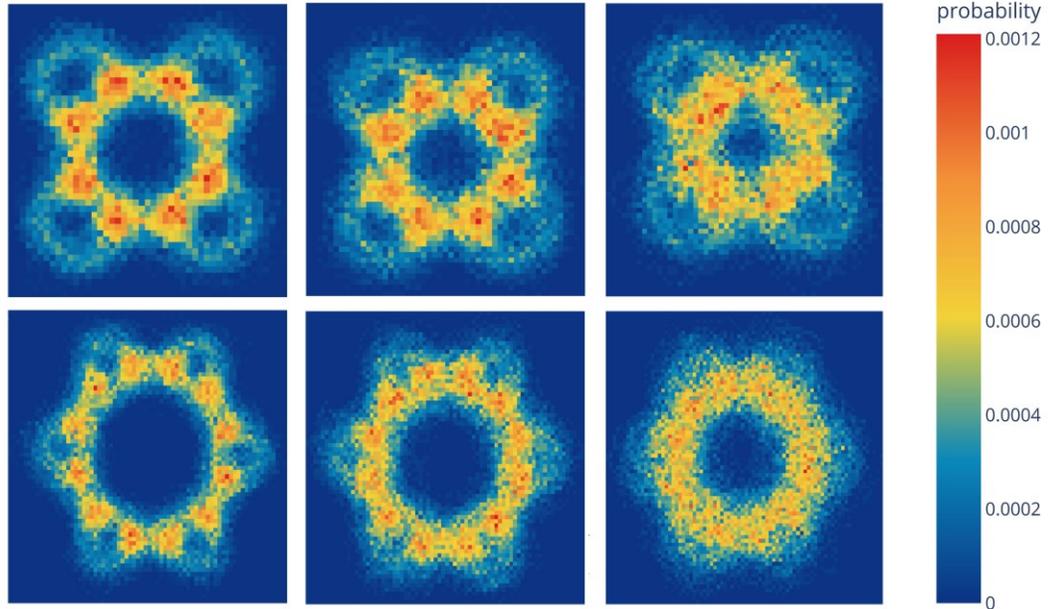

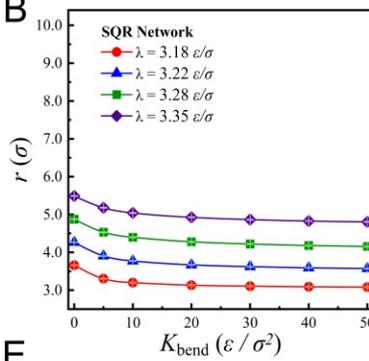 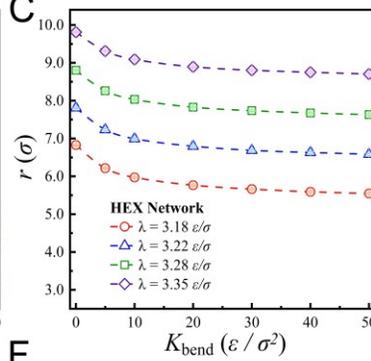 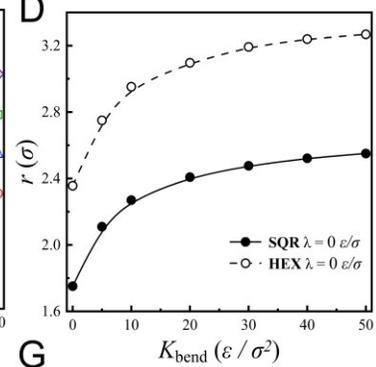

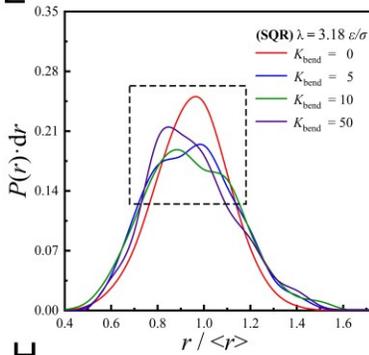 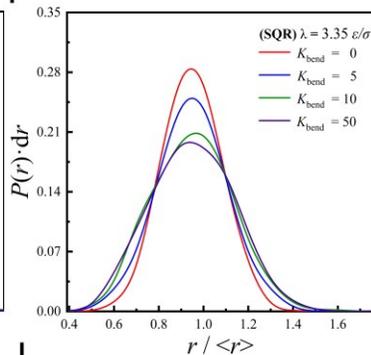 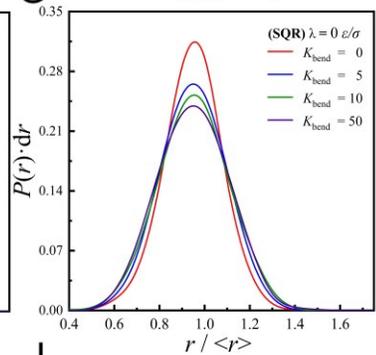

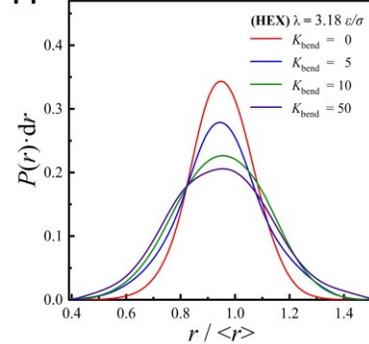 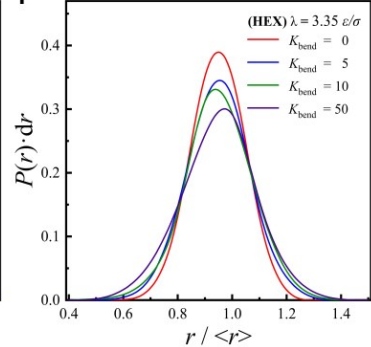 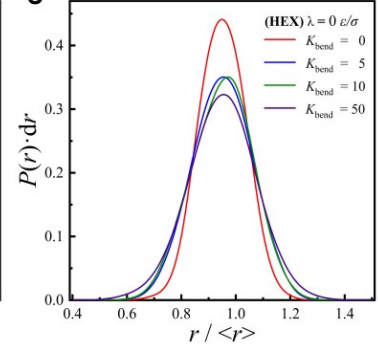

**Figure 2. (A)** Particle density heatmap of pores in the SQR and HEX network. **(B) and (C).** Curves of pore size $r$ as a function of $K_{bend}$ ranging from 0 to 50 $\varepsilon/\sigma^2$ under $\lambda$ = 3.18 $\varepsilon/\sigma$, 3.22 $\varepsilon/\sigma$, 3.28 $\varepsilon/\sigma$ and 3.35 $\varepsilon/\sigma$ in SQR network and HEX network. To enable a direct comparison of pore size between the two networks, the spatial range of the Y-axis has been normalized to the same scale. **(D)** Pore size curves of SQR and HEX networks with $K_{bend}$ under tension-free conditions. **(E) to (J).** Pore size probability distribution curves for the SQR network with $K_{bend}$ = 0, 5, 10, 50 $\varepsilon/\sigma^2$ but **(E)** $\lambda$ = 3.18 $\varepsilon/\sigma$, **(F)** $\lambda$ = 3.35 $\varepsilon/\sigma$ and **(G)** $\lambda$ = 0 $\varepsilon/\sigma$. The **(H)** $\lambda$ = 3.18 $\varepsilon/\sigma$, **(I)** $\lambda$ = 3.35 $\varepsilon/\sigma$ and **(J)** $\lambda$ = 0 $\varepsilon/\sigma$. are pore size probability distribution curves for the HEX network with $K_{bend}$ = 0, 5, 10, 50 $\varepsilon/\sigma^2$. Black dashed rectangle in **(E)** highlights the splitting and broadening of the distribution curves.

To further investigate the characteristics of pore size ($r$), we plotted the pore size probability distribution curves. In Fig. 2(E) to (J), the Y-axis represents the probability $P(r)\,dr$, while the X-axis shows the normalized ratio $r/\langle r \rangle$, where $\langle r \rangle$ is the mean *r*. To reduce overlapping of distribution curves, only systems with different topological tension strengths and polymer chain stiffness values of $K_{bend}$ = 0, 5, 10, 50 $\varepsilon/\sigma^2$ were selected for visualization. Fig. 2E corresponds to the pore size distribution of the SQR network under a topological tension strength of $\lambda$ = 3.18 $\varepsilon/\sigma$ In the fully flexible polymer chain system ($K_{bend}$ = 0), the distribution exhibits a single peak centered around $r/\langle r \rangle$=1.0. However, as chain stiffness increases, the distribution shape changes. At $K_{bend}$ = 5 $\varepsilon/\sigma^2$, the distribution broadens and splits into a bimodal form, with the highest peak still locating near $r/\langle r \rangle$ = 1.0, while a secondary, smaller peak appears at around $r/\langle r \rangle$ = 0.8. For $K_{bend} \geqslant$ 10 $\varepsilon/\sigma^2$, the distribution shifts, and the peak at $r/\langle r \rangle$ = 0.8 becomes the dominant one. As chain stiffness further increases, the distribution evolves back into a single-peak shape. This behavior indicates that *r* has two characteristic sizes, representing larger and smaller pore radius. Fig. 2F shows the pore size distribution for the SQR network under the strongest topological tension ($\lambda$ = 3.35 $\varepsilon/\sigma$). Compared to Fig. 2E, it is evident that with increasing chain stiffness, the primary peak of the distribution consistently remains centered at $r/\langle r \rangle$ = 1.0, and no splitting or shifting occurs. Instead, the distribution exhibits only minor broadening. In addition, we plotted the pore-size distribution of the tension-free SQR network, as shown in Fig. 2G. The distribution curves at different backbone stiffness values are similar to those in Fig. 2F, indicating that under tension-free conditions the $r$ also exhibits only a single characteristic value. The stronger topological tension eliminates the smaller characteristic size of *r*, while the broadening of the distribution persists, suggesting that the smaller characteristic size of *r* is related to $\lambda$. The mechanisms behind these observations can be attributed to two factors. Firstly, polymer chain stiffness controls the volume of the polymer rings, which in turn regulates *r*, corresponding to one characteristic size in the distribution curves. Besides, based on the catenation structures shown in Fig. 1, we hypothesize that changes in topological tension strength affect the rotational freedom of the polymer rings. At high chain stiffness, the rotational motion of larger polymer rings occupies additional pore volume, corresponding to the smaller characteristic size in the distribution curves.

Additionally, Fig. 2H, Fig. 2I and Fig. 2J show the pore size probability distribution for the HEX network under $K_{bend}$ = 0, 5, 10, 50 $\varepsilon/\sigma^2$ and $\lambda$ = 3.18 $\varepsilon/\sigma$, 3.35 $\varepsilon/\sigma$ respectively. In both cases, the primary peaks of the distribution curves are consistently centered around $r/\langle r \rangle$ = 0.9. While changes in polymer chain stiffness result in significant broadening of the distribution curves, where the phenomena of peak splitting and shifting are not evident. Notably, in Fig. 2I, an increase in topological tension strength further reduces the broadening of the distribution curves. This is consistent with the results observed in Fig. 2E and Fig. 2F.

We suggest that the two regulatory mechanisms proposed above remain applicable in the HEX network. However, differences in the unit cell arrangement of the network's pores lead to variations in the distribution curve shapes compared to the SQR network. Based on the Maxwell relation and network topology, the SQR network is a typical isostatic network, while the HEX network is hypostatic. In both networks, the same catenane connection between polymer rings provides the connection points with translational and rotational degrees of freedom without using energy. In the hypostatic structure of the HEX network, extra zero modes exist. The polymer rings corresponding to these zero modes do not change their elastic potential energy when they move. If the change in the stiffness of the polymer chains is viewed as a perturbation, the extra zero modes in the HEX network supply energy redundancy that eliminates this perturbation. This results in distribution curves that are difficult to distinguish. We will discuss this point in detail in the section on the correlation between adjacent pores later.

To further detail the mechanisms by which polymer ring conformational fluctuations and rotations regulate pore size ($r$) under the topological feature of catenane structures, and to validate the proposed mechanism that the r is co-regulated by the conformational and rotational degrees of freedom of the polymer rings—thereby giving rise to two characteristic length scales, we first calculated the radius of gyration ($R_g$) to characterize the conformation of polymer chains. In our simulated systems, the SQR and HEX networks consist of polymer rings with two different valences, i.e., where $v$ represents the number of rings connected to a given ring through catenane structures. In the SQR network, there are two types of rings with $v = 4$ and $v = 2$, while in the HEX network, we have $v = 3$ and $v = 2$. Additionally, we computed the $R_g$ of free polymer rings with $v = 0$ for comparison to highlight the effect of catenane topological structures on conformational fluctuations. In Fig. 3A, we present the variation in the $R_g$ of rings with different degrees as a function of chain stiffness ($K_{bend}$) under the topological tension strength $\lambda = 3.18\ \varepsilon/\sigma$. For rings with different degrees, $R_g$ increases with $K_{bend}$, and this increasing trend levels off when $K_{bend} \geqslant 30\ \varepsilon/\sigma^2$. This observation indicates that increased chain stiffness leads to larger ring volumes, providing preliminary evidence that the reduction in r with increasing chain stiffness originates from the occupation of pore space by polymer rings. At the same chain stiffness, comparing rings with different degrees reveals a proportional relationship between $R_g$ and v. Namely, rings with higher v exhibit larger $R_g$, with the difference in $R_g$ between rings of varying v being largest for fully flexible chains ($K_{bend} = 0$) and gradually decreasing as chain stiffness increases. When $K_{bend} \geqslant 30\ \varepsilon/\sigma^2$, the $R_g$ curves for rings with different v nearly overlap, suggesting that at high chain stiffness, polymer rings of different degrees possess similar molecular volumes. We attribute the observed expansion of ring conformations and the reduction in expansion differences with increasing chain stiffness to the unique topology of catenane structures.

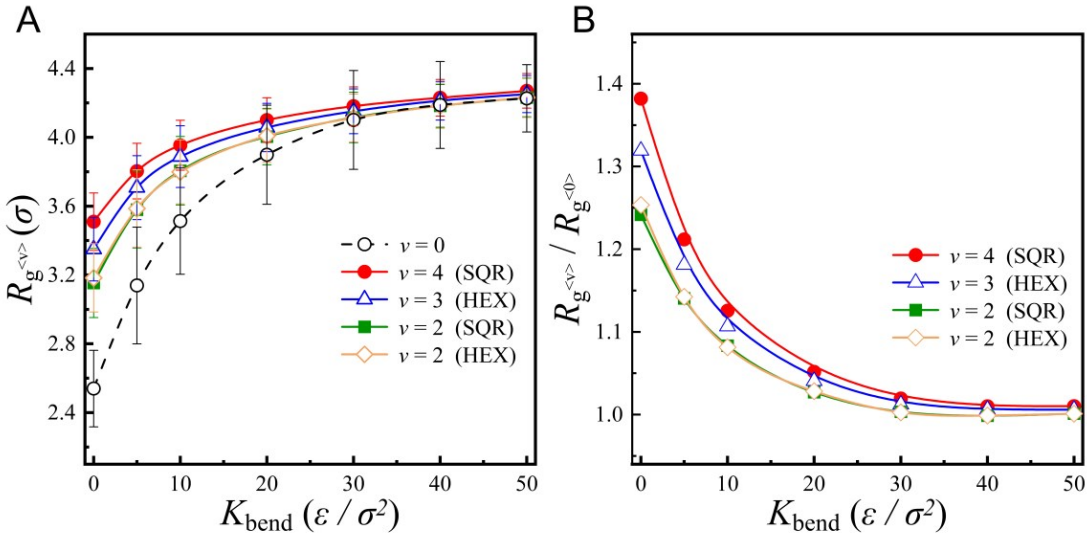

**Figure 3.** **(A)** Radius of gyration ($R_g$) of polymer rings as a function of $K_{bend}$ under $\lambda = 3.18\ \varepsilon/\sigma$. The curves correspond to rings with $v= 2, 3, 4$ from the SQR and HEX networks. The $R_g$ curve for free polymer rings ($v= 0$) is marked with a black dashed line. **(B)** Expansion coefficient $R_g^{<v>}/R_g^{<0>}$ as a function of $K_{bend}$

To quantify the above effect, we defined a normalized expansion coefficient, calculated as $R_g^{<v>}/R_g^{<0>}$, where $R_g^{<v>}$ represents the $R_g$ of rings with v > 0, and $R_g^{<0>}$ represents the $R_g$ of free rings ($v = 0$). The magnitude of the value $R_g^{<v>}/R_g^{<0>}$ directly quantifies the effect of the catenane topology on ring conformation. When the value is close to 1.0, the lower additional expansion of the polymer ring conformation is induced by the catenane topology.

In Fig. 3B with the relationship of $R_g^{<v>}/R_g^{<0>}$ vs. $K_{bend}$, at low chain stiffness, the four curves are arranged sequentially from top to bottom, with $R_g^{<v>}/R_g^{<0>}$ values exceeding 1.2. This indicates that the catenane structure induces additional conformational expansion, and rings with higher v exhibit greater expansion due to the influence of the catenane topology. As chain stiffness increases, $R_g^{<v>}/R_g^{<0>}$ decrease, converging to 1.0 when $K_{bend} \geqslant 30\ \varepsilon/\sigma^2$. This indicates that at high chain stiffness, the catenane topology has negligible impact on ring conformations. The additional swelling induced by catenation stems from the way this topology fills the interior of a polymer ring through an excluded-volume effect. At low chain stiffness, an isolated ring polymer prefers a crumpled conformation with a small $R_g$. When two rings are catenated, the penetrating chain segments occupy the ring's interior, introducing an excluded-volume penalty that drives the ring away from its most compact state and therefore produces extra expansion. As the chain stiffness increases, the ring's $R_g$ also grows and the molecule adopts a more extended conformation; the enlarged internal cavity can now accommodate the interpenetrating segments of the catenated partner, rendering the catenation-induced excluded-volume effect almost negligible. Consequently, at high chain stiffness, the expansion coefficients for all curves converge to 1.0.

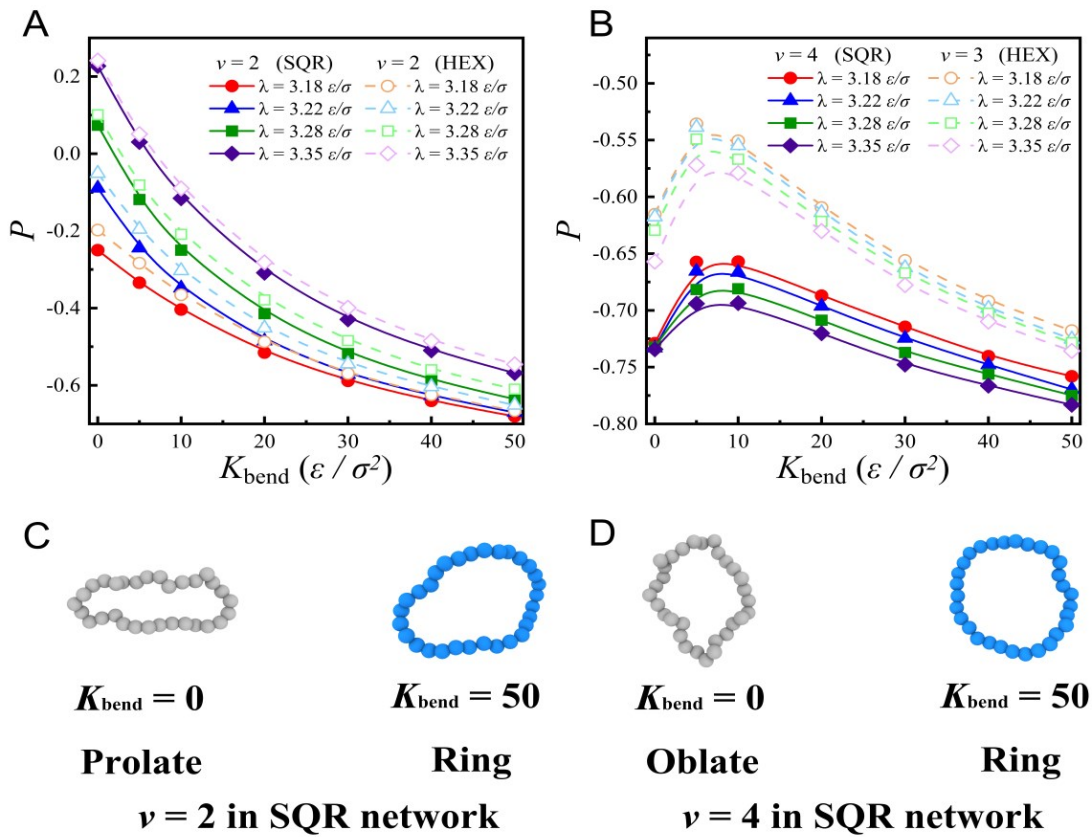

**Figure 4.** **(A)** Variation of $P$ with $K_{bend}$ for $v = 2$ polymer rings in SQR and HEX networks under $\lambda$=3.18 $\varepsilon/\sigma$, 3.22 $\varepsilon/\sigma$, 3.28 $\varepsilon/\sigma$ and 3.35 $\varepsilon/\sigma$. **(B)** Variation of $P$ with Kbend for $v = 4$ polymer rings in the SQR network and $v$ =3 polymer rings in the HEX network under $\lambda$=3.18 $\varepsilon/\sigma$, 3.22 $\varepsilon/\sigma$, 3.28 $\varepsilon/\sigma$ and 3.35 $\varepsilon/\sigma$. **(C)** Schematic of polymer ring conformation transitions: a $v = 2$ ring transitions from a prolate to a ring conformation. **(D)** A $v = 4$ ring transitions from an oblate to a ring conformation.

We further used the Prolateness ($P$) to investigate the complex relationship between the catenane topology, chain stiffness, and the conformations of polymer rings, aiming to clarify how ring conformations regulate the pore size ($r$). Prolateness is a quantitative metric, similar to the sphericity factor, that describes the transition of polymer ring conformations between ideal circular and cylindrical shapes. It is calculated from the gyration tensor of the ring polymer as follows:

$$P = \frac{(2\lambda_1 - \lambda_2 - \lambda_3)(2\lambda_2 - \lambda_1 - \lambda_3)(2\lambda_3 - \lambda_1 - \lambda_2)}{2(\lambda_1^2 + \lambda_2^2 + \lambda_3^2 - \lambda_1\lambda_2 - \lambda_1\lambda_3 - \lambda_2\lambda_3)^{3/2}} . \quad (7)$$

In Equation 7, $\lambda 1, \lambda 2, \lambda 3$ are the eigenvalues of the gyration tensor, ranked in descending order as $\lambda 1 \geqslant \lambda 2 \geqslant \lambda 3$. The Prolateness ranges from $-1 \leqslant P \leqslant 1$: $P = -1$ indicates an oblate conformation resembling an ideal circular ring, $P = 1$ corresponds to a prolate (elongated rod-like) conformation, and $P \approx 0$ suggests a collapsed structure. Fig. 4A illustrates the variation of $P$ for $v = 2$ polymer rings in SQR and HEX networks as a function of chain stiffness under different topological tension strengths. As chain stiffness increases, the $P$ values decrease, indicating that the ring conformations transition toward more extended oblate shapes with increasing stiffness. Under low chain stiffness, comparison of *P* of polymer rings among different topological tension strengths reveals that *P* increases from approximately -0.2 to 0.2 as the topological tension strength increases. This suggests that stronger topological tensions promote a transition toward prolate conformations under low chain stiffness. However, under high chain stiffness, the differences in *P* values across topological tension strengths diminish to around 0.1 (ranging from -0.65 to -0.55), indicating that the additional effects of topological tensions on ring conformations are primarily significant under low chain stiffness. Fig. 4B presents the variation of *P* for $v = 4$ polymer rings in the SQR network and $v = 3$ polymer rings in the HEX network under different topological tension strengths. The *P* values remain below -0.5 and exhibit a unique trend of first increasing and then decreasing as chain stiffness grows. This behavior indicates that rings with higher degrees in both networks adopt predominantly oblate conformations but show slight collapsing tendencies at intermediate chain stiffness. Stronger topological tensions further drive the rings toward oblate conformations, reducing the collapsing tendency observed at intermediate chain stiffness.

Correlating the ring polymer's radius of gyration ($R_g$) with its cylindricity (*P*) reveals an extra deformation of the ring conformation in the flexible-chain regime of low bending stiffness. Returning to Fig. 2E, one sees that for $K_{bend} < 10$ the bimodal pore-size distribution is dominated by the larger characteristic value $r/\langle r \rangle = 1.0$. We propose that the appearance of this peak originates from additional conformational perturbations imposed by the catenane topology under flexible-chain conditions, and that these perturbations are fundamentally entropic. This claim can be substantiated by a simple derivation: the entropic (elastic) force of a polymer chain is obtained by taking the partial derivative of the Helmholtz free energy *A* with respect to the end-to-end distance R:

$$F = \frac{\partial A}{\partial R} . \tag{8}$$

For a polymer ring, which has no free chain ends, the $R_g$ can be used in place of the end-to-end distance. Expanding the Helmholtz free energy *A* then gives:

$$F = \frac{\partial A}{\partial R_g} = \frac{\partial (U - TS)}{\partial R_g} . \tag{9}$$

Accordingly, the elastic modulus E of a polymer ring can be obtained by taking the second partial derivative of the Helmholtz free energy *A* with respect to the $R_g$:

$$E = \frac{\partial F}{\partial R_g} = \frac{\partial^2 A}{\partial R_g^2} . \tag{10}$$

It is well established that increasing the bending stiffness of a polymer chain reduces its conformational entropy. Within the flexible-chain regime ($K_{bend} < 10$), we further assume that the system's internal energy *U* varies negligibly with stiffness. Under these conditions, the expression simplifies to:

$$F = \frac{-T \partial S}{\partial R_g} . \tag{11}$$

Taken together with Fig. 3A, which shows that the ring's $R_g$ grows as the bending stiffness increases, we infer that the ring's elastic modulus likewise rises with $K_{bend}$. The ring therefore becomes intrinsically stiffer, and a larger external force is required to perturb its conformation.

Building on the above discussion, we now introduce the influence of catenation topology. In the simulations, ring polymers are interlocked to form a network that supports a topological tension ($\lambda$). Fig. 3A shows that, at low bending stiffness, the $R_g$ of a non-catenated ring is markedly smaller than that of a catenated ring. A catenated ring therefore has an entropic incentive to contract and reduce its $R_g$. Because the rings are mutually interlocked and the chains are impenetrable, the contraction of one ring necessarily exerts a pulling force on its partner; the elastic forces arising from these reciprocal tugs constitute the network's topological tension and are fundamentally entropic in origin. According to the modulus analysis, flexible rings possess a lower elastic modulus and hence deform more readily under this entropic pull, which explains the pronounced prolate and oblate conformations observed for low-stiffness rings in Fig. 4A and Fig. 4B. Such deformations shrink the area occupied by each ring and, in turn, enlarge the surrounding pore, yielding the dominant peak at $r/\langle r \rangle = 1.0$ in the pore-size probability distribution. As the network's topological tension λ increases, this entropic effect becomes predominant, causing the smaller characteristic peak to disappear from the distributions in Fig. 2E and 2H.

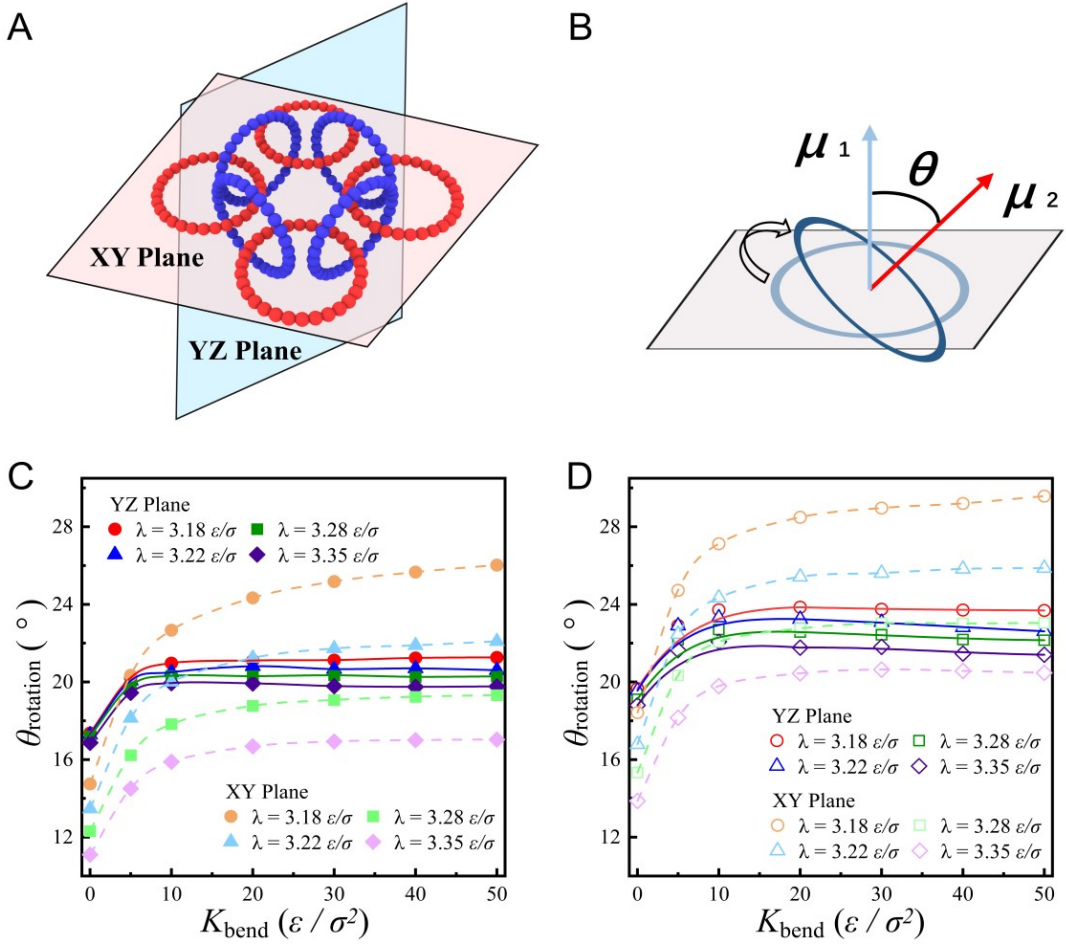

**Figure 5. (A) and (B)** Schematic representation of the rotational angle between a polymer ring and the reference plane, defined by the angle between the ring's normal vector and the plane normal. **(C)** Curves showing the changes in rotational angles of two types of polymer rings in the SQR network. **(D)** Curves showing the changes in rotational angles of two types of polymer rings in the HEX network.

In the low–stiffness regime, conformational-entropy–dominated regulation of pore size accounts only for the larger peak in the distribution; the appearance of the smaller characteristic value at $r/\langle r \rangle = 0.8$ requires a different mechanism. Building on our earlier conjecture, we ascribe this smaller peak to the rotational degrees of freedom of the rings. To test this idea, we quantified the average rotation angle ($\theta$) of rings with connectivity $v = 2, 4$ in the square (SQR) network and $v = 2, 3$ in the hexagonal (HEX) network. A cautionary note is necessary: the "standard plane'' from which rotations are measured differs among topologies. Fig. 5A sketches the standard planes in the SQR network. For $v = 4$ rings the initial plane is the un-relaxed XY-plane (highlighted in red), whereas for $v = 2$ rings the two initial planes lie parallel to the Y and X-axes, respectively (Y-axes is shown in blue). By analogy, the HEX network employs analogous reference planes, though the $v = 2$ rings there possess a richer set of initial planes. As depicted in Fig. 5B, we determine the $\theta$ by taking the angle between the normal vector of the ring's instantaneous largest-area plane and that of its initial plane throughout the trajectory. Fig. 5C presents the $\theta$ for rings in the SQR network. The angle grows with bending stiffness and saturates for $K_{bend} \geqslant 10\ \varepsilon/\sigma^2$. At fixed $K_{bend}$, raising the topological tension $\lambda$ reduces the $\theta$; the drop is markedly stronger for $v = 4$ rings than for $v = 2$ rings as $\lambda$ increases from 3.18 $\varepsilon/\sigma$ to 3.35 $\varepsilon/\sigma$. Fig. 5D shows analogous trends for the HEX network. A comparison reveals that, overall, rings in the HEX network rotate through larger angles than their SQR counterparts.

These observations indicate that, as stiffness increases, conformational entropy decreases and the bending energy becomes dominant. Rings then adopt an "ideal'' conformation with a larger internal cavity that can accommodate interpenetrating chains; the reduced pulling among rings at the same $\lambda$ endows the system with additional rotational freedom, explaining the monotonic growth and eventual saturation of the rotation angle (θ), in line with the $R_g$ trends in Fig. 3A. Increasing λ, however, strengthens the mutual tethers between rings and suppresses their rotational freedom. Because this constraint scales with connectivity ν, rings with higher ν exhibit both smaller absolute θ and larger decrements when λ is raised, consistent with the sharper decline observed for the $ν=4$ rings in the SQR network and the generally larger angles in the HEX network.

An ideal-ring conformation, with its larger $R_g$, occupies more of the pore volume and thereby reduces the pore size. The additional rotational freedom allows the ring to intrude further into the pore cross-section, especially for $ν = 2$ rings, whose rotation promotes deeper penetration into the projection plane (the XY-plane). This extra occupancy mechanism produces the smaller peak at $r/\langle r \rangle = 0.8$ in the pore-size probability distribution. Enhancing λ curtails the rotation, suppressing this additional occupancy and accounting for the disappearance of the 0.8 peak and the sharpening of the distribution curves in Fig. 2F and 2I.

In summary, by using the ring polymer's radius of gyration ($R_g$), prolateness (P), and rotation angle (θ), we have systematically investigated two distinct mechanisms that regulate pore size in SQR and HEX networks: the entropic elasticity of the polymer chains and the rotational motion of the ring polymers. These two mechanisms counteract each other, giving rise to the splitting of the pore-size probability distribution curve within specific regions of parameter space. Under low chain-stiffness conditions, pore size is governed predominantly by the chain's entropic elasticity; this effect enlarges the pore-size radius, producing larger characteristic pores. Increasing the topological tension (λ) extends the range of chain-stiffness parameters over which this entropic mechanism dominates, leading to the disappearance of the smaller-size peak in the distribution curve. At high chain stiffness, the rotational motion of the ring polymers becomes dominant, introducing an additional reduction in the pore-size radius, which appears as the lower characteristic peak in the distribution. Within intermediate parameter ranges, the two effects compete, so the two characteristic pore sizes exhibit different statistical probabilities.

## Correlation and Network Structure

Next, we discuss the overall properties of SQR and HEX networks, including the interaction between pores and the structure of the networks. First, we calculated the spatial correlation function for the pore size r in both networks by Equation 12:

$$C(i) = \frac{1}{N} \sum_{x=1}^{N} \frac{\langle [r(i) - \langle r \rangle][r(i+i') - \langle r \rangle] \rangle}{\langle [r(i) - \langle r \rangle]^2 \rangle} . \qquad (12)$$

In Eq. 12 the quantity $C(i)$ denotes the value of the spatial correlation function. Here the separation between pores is expressed in terms of their index difference rather than an absolute metric distance. The symbol $r(i)$ represents the pore-size radius of the pore with index $i$. Specifically, the integer $i$ designates the correlation evaluated between pores whose indices differ by $i = 1$ corresponds to two adjacent pores, $i = 2$ to pores with one pore in between, and so forth. The variable $N$ is the total number of pores in the network. Fig. 6A represents the SQR network. Under strong topological tensions (λ = 0.467) and low $K_{bend}$, there is a certain negative correlation between adjacent pore sizes, where the size of one pore tends to reduce the size of neighboring pores. As $K_{bend}$ increases, this negative correlation diminishes rapidly, and no significant correlation exists between adjacent pores. Similarly, a decrease in λ also leads to the rapid disappearance of negative correlation. For the HEX network, as shown in Fig. 6B, the results reveal a very weak or negligible positive correlation between adjacent pores as $K_{bend}$ increases. Changes in the topological tension strength λ also fail to enhance this correlation.

This phenomenon comes from the combined effect of the catenane topology tension and the lattice structure that forms the network unit. As discussed in Fig. 2, the SQR network can be regarded as an approximately isostatic network and the HEX network as an approximately hypostatic network. Previous studies have shown that an isostatic network is in a critical state. Zero modes exist at the boundaries of the lattice, and small perturbations can propagate along these boundaries. We further propose that the pulling effect introduced earlier dictates how closely our Olympic network resembles an ideal point-spring cross-linked network; the stronger the pulling effect, the tighter the conformational constraints and interactions between polymer chains. When the chain stiffness is low—so conformational entropy dominates—and the topological tension is high, the pulling effect becomes pronounced, making the catenane connections between rings behave almost like permanent cross-link points. Based on this assumption, we can explain the phenomenon shown in Fig. 6A. When $K_{bend}$ is low and λ is high, the strong tension effect between polymer rings makes the network approximately equivalent to an ideal isostatic network. In this case, fluctuations in the size of a single pore can be transmitted to adjacent pores. This is shown by the negative correlation of adjacent pore sizes indicated by the yellow dots in Fig. 6A. An increase in the size of one pore leads to a decrease in the size of adjacent pores and vice versa. As the $K_{bend}$ increases and the $\lambda$ decreases, the tension effect weakens. This results in extra degrees of freedom for the polymer rings due to the catenane topology, which in turn weakens the influence of pore size fluctuations on adjacent pores. This is reflected by the rapid decrease of the adjacent pore correlation function value toward zero when the $K_{bend}$ increases and the $\lambda$ decreases. In contrast, in a hypostatic network, additional zero modes exist within the lattice. Fluctuations in pore size tend to cause deformation of the lattice itself, which makes it difficult to transmit perturbations over longer distances. This is shown in Fig. 6B, where even under the strongest topological tension effect, there is almost no correlation between adjacent pores. This phenomenon indicates that the catenation can act as an extra topological factor to introduce more topological flexibility into polymer networks and demonstrates the unique nature of the catenane topology.

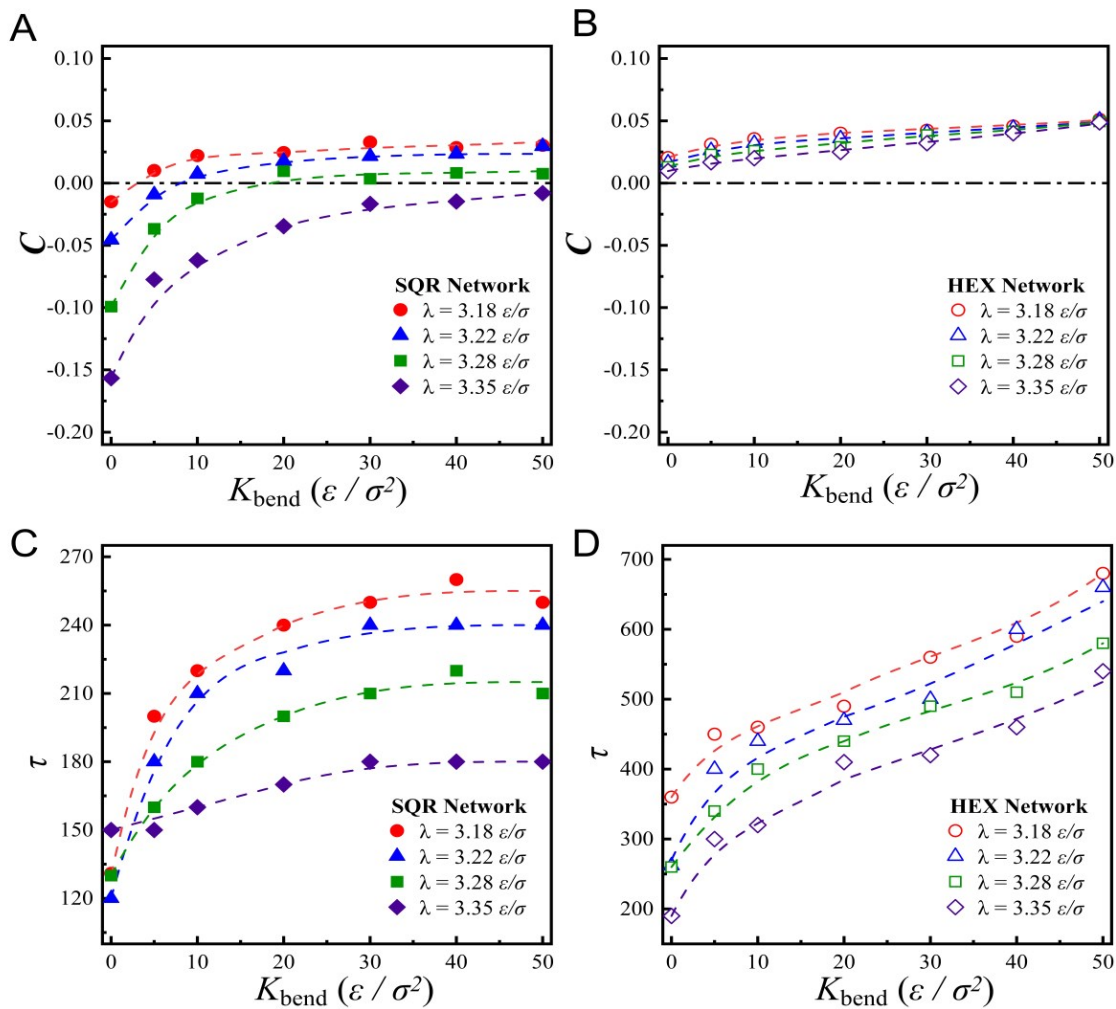

**Figure 6.** **(A)** Variation of the adjacent pore size correlation function $C(i)$ with $K_{bend}$ for the SQR network under parameters $\lambda$=3.18 $\varepsilon/\sigma$, 3.22 $\varepsilon/\sigma$, 3.28 $\varepsilon/\sigma$ and 3.35 $\varepsilon/\sigma$. **(B)** Variation of the adjacent pore size correlation function $C(i)$ with $K_{bend}$ for the HEX network under same tension parameters. **(C)** Variation of pore size autocorrelation time $\tau(t)$ with $K_{bend}$ for the SQR network under parameters $\lambda$=3.18 $\varepsilon/\sigma$, 3.22 $\varepsilon/\sigma$, 3.28 $\varepsilon/\sigma$ and 3.35 $\varepsilon/\sigma$. **(D)** Variation of pore size autocorrelation time $\tau(t)$ with $K_{bend}$ for the HEX network under same tension parameters.

We have also evaluated the pore self-correlation time, $\tau$, across the different regions of parameter space. The quantity $\tau$ is defined in Equation. 13.

$$\tau(t) = \frac{1}{N}\sum_{i=1}^{N} \frac{\langle[r(t)-\langle r\rangle][r(t+t')-\langle r\rangle]\rangle}{\langle[r(t)-\langle r\rangle]^2\rangle}. \tag{13}$$

where $\tau(t)$ is the pore size (r) of a given pore at time $t$, and $N$ is the total number of pores in the network. Fig. 6C and 6D illustrate the variation of pore autocorrelation time with chain stiffness $K_{\text{bend}}$ and topological tension strength $\lambda$. The self-correlation time quantifies the extent to which the present pore size remembers its past values. We define the self-correlation time $\tau$ as the earliest time at which the normalized autocorrelation function $\tau(t)$ falls below 0.1; once $\tau(t) < 0.1$, the pore size is deemed statistically independent of its previous state. In Fig. 6C with the autocorrelation time curves for the SQR network, an increase in $K_{\text{bend}}$ leads to a gradual rise in autocorrelation time, while an increase in $\lambda$ results in a decrease in autocorrelation time. Similarly, Fig. 6D presents the autocorrelation time curves for the HEX network, demonstrating the same trend. We attribute this consistent behavior to the fact that the polymer rings in both networks are connected through the same catenane topology. At lower $K_{\text{bend}}$ and higher $\lambda$, the polymer rings behave similarly to "crosslinked" structures, where conformational fluctuations between rings easily influence each other, resulting in shorter autocorrelation times. Conversely, higher $K_{\text{bend}}$ and lower $\lambda$ create space that allows for a certain degree of free movement between polymer rings, making it difficult for conformational fluctuations of individual rings to propagate, thereby leading to longer autocorrelation times.

Through a systematic analysis of the spatial correlation function of pore size $C(i)$, and the temporal self-correlation time $\tau(t)$, we uncover the correlation characteristics of SQR and HEX networks—two distinct topological architectures—across different chain bending stiffnesses $K_{\text{bend}}$ and topological tensions $\lambda$. Spatially, the SQR network exhibits pronounced negative nearest-neighbor correlations in the low-$K_{\text{bend}}$, high-$\lambda$ regime, whereas the HEX network shows almost none, underscoring the fundamental difference in disturbance-propagation capability between isostatic and hypostatic lattices. Temporally, both networks display longer self-correlation times as $K_{\text{bend}}$ increases, indicating that, under varying conditions, the catenane topology can act either as a "quasi-cross-link," suppressing pore-size fluctuations, or, once relaxed, as a source of additional motional freedom that modulates pore size. Structurally, the influence of catenation manifests only in local adjustments of pore-size dimensions without perturbing the network's long-range order.

## 4. CONCLUSION

In living systems, assemblies of interlocked rings—such as catenated mitochondrial/plasmid DNA, kDNA sheets, and capsid-like ring lattices, which gate molecular access through ring-defined pores. The efficiencies of diffusion-limited steps (e.g., replication, editing, enzyme binding or reaction, and genome release) depend on what pore sizes are available over space and time. Here we used a two-dimensional periodic-boundary (PBC) model to study the bulk of catenated networks and to separate universal effects from system-specific chemistry and edges.

The results show that in both SQR and HEX networks, as $K_{\text{bend}}$ increases from 0 to 50, the average pore radius r decreases, but the rate of decrease levels off when $K_{\text{bend}} > 20\ \varepsilon/\sigma^2$. This phenomenon can be attributed to conformational transitions of polymer rings. At low chain stiffness, ring molecules adopt collapsed conformations, occupying smaller volumes within the pores. As chain stiffness increases, the rings gradually stretch, occupying larger spatial volumes and thereby reducing the effective pore size.

The topological tension strength $\lambda$ directly modulates the spatial arrangement of polymer rings, thus influencing the pore size. At fixed $K_{\text{bend}}$, increasing $\lambda$ enlarges the pore size in both SQR and HEX networks, as greater distances between rings create larger void spaces in the network.

An important observation is the emergence of bimodal pore size distributions under moderate $K_{\text{bend}}$ values, particularly at low $\lambda$ in the SQR network. This phenomenon arises from the coexistence of two competing mechanisms: the entropic elasticity of polymer rings and their rotational degrees of freedom. Rings with low stiffness exhibit lower entropic elasticity and are easily deformed under the pulling effects of connecting rings, resulting in elongated conformations that occupy smaller pore spaces, corresponding to the larger peak in the bimodal distribution. At intermediate chain stiffness, rings adopt extended conformations, and their increased rotational degrees of freedom further reduce the effective pore size, contributing to the smaller peak in the distribution. Under high $\lambda$, the bimodal distribution is suppressed, indicating that stronger topological tensions limit the rotational freedom of polymer rings, thereby homogenizing the pore size distribution.

Analysis of adjacent pore size correlations revealed significant differences between SQR and HEX networks. In the SQR network, adjacent pores exhibit negative correlations under low $K_{\text{bend}}$ and high $\lambda$, indicating that larger pore sizes restrict the size of neighboring pores. This effect diminishes as $K_{bend}$ increases or λ decreases, reflecting a transition from tightly constrained to more flexible network structures. In contrast, adjacent pore sizes in the HEX network show almost no correlation across the entire parameter range. Pore size autocorrelation time τ further highlights the dynamic behavior of the networks. At high $K_{bend}$ and low λ, both SQR and HEX networks exhibit longer τ, indicating more persistent pore size fluctuations in less constrained networks. This result underscores the roles of polymer chain stiffness and topological tensions in governing network dynamics.

The differences between SQR and HEX networks underscore the role of network geometry in regulating topological effects. As an isostatic network, the SQR network maintains a critical balance between its degrees of freedom and its state of self stress, making it more sensitive to changes in Kbend and λ. In contrast, the hypostatic HEX network has extra degrees of freedom due to its underconstrained nature and shows lower sensitivity to topological tensions. This difference is evident in the correlation of pore sizes, the propagation of disturbances within the network, and the radial distribution function of pore centroids.

In summary, we used a 2D periodic model to study bulk behavior in catenated ring networks and asked how ring stiffness ($K_{bend}$) and topological tension (λ) shape pore statistics that be transport. These results yield testable links to FRAP, SMT and single-particle tracking readouts of diffusion and enzyme access, providing a biophysics baseline for pore-controlled steps in replication, editing, and genome release. Our framework is system-agnostic—it does not model kDNA, HK97, or any specific organism, but offers predictions for interior transport that experiments can evaluate across diverse catenated assemblies. We focused on the bulk limit under PBCs and did not include edges, strand passage, or sequence/protein chemistry; extending the model to these factors is a natural next step. By connecting network topology and mechanics to accessible pore sizes and their dynamics, this work links ring architecture to diffusion-limited biochemistry in living systems.

## ASSOCIATED CONTENT

## AUTHOR INFORMATION

### Corresponding Author

hongliu@m.scnu.edu.cn (H. L.).

guojie.zhang@gzhu.edu.cn. (G. Z.);

**Notes**

The authors declare no competing financial interests.

## ACKNOWLEDGMENT

This work is supported by the National Natural Science Foundation of China (22373024, 22273027), the Natural Science Foundation of Guangdong Province (2025A1515010731) and the Open Research Fund of Songshan Lake Materials Laboratory (2023SLABFK11). Hong Liu gratefully acknowledges the support from the Alexander von Humboldt Foundation.

# REFERENCES

(1) Hotta, Y.; Bassel, A. MOLECULAR SIZE AND CIRCULARITY OF DNA IN CELLS OF MAMMALS AND HIGHER PLANTS. *Proc. Natl. Acad. Sci. U.S.A.* **1965**, *53* (2), 356–362. https://doi.org/10.1073/pnas.53.2.356.

(2) Gaubatz, J. W. Extrachromosomal Circular DNAs and Genomic Sequence Plasticity in Eukaryotic Cells. *Mutation Research/DNAging* **1990**, *237* (5–6), 271–292. https://doi.org/10.1016/0921-8734(90)90009-G.

(3) Shapiro, T. A. Kinetoplast DNA Maxicircles: Networks within Networks. *Proc. Natl. Acad. Sci. U.S.A.* **1993**, *90* (16), 7809–7813. https://doi.org/10.1073/pnas.90.16.7809.

(4) Chiarantoni, P.; Micheletti, C. Effect of Ring Rigidity on the Statics and Dynamics of Linear Catenanes. *Macromolecules* **2022**, *55* (11), 4523–4532. https://doi.org/10.1021/acs.macromol.1c02542.

(5) Rawdon, E. J.; Kern, J. C.; Piatek, M.; Plunkett, P.; Stasiak, A.; Millett, K. C. Effect of Knotting on the Shape of Polymers. *Macromolecules* **2008**, *41* (21), 8281–8287. https://doi.org/10.1021/ma801389c.

(6) Wu, Q.; Rauscher, P. M.; Lang, X.; Wojtecki, R. J.; De Pablo, J. J.; Hore, M. J. A.; Rowan, S. J. Poly[ *n* ]Catenanes: Synthesis of Molecular Interlocked Chains. *Science* **2017**, *358* (6369), 1434–1439. https://doi.org/10.1126/science.aap7675.

(7) Xie, X.; Wu, L.; Sun, H.; Yan, X.; Zhu, X. Phase Behavior and Liquid Crystalline Ordering of [2]Catenated Molecular Systems. *Macromolecules* **2023**, *56* (16), 6189–6198. https://doi.org/10.1021/acs.macromol.3c00186.

(8) Takashima, R.; Aoki, D.; Otsuka, H. Synthetic Strategy for Mechanically Interlocked Cyclic Polymers via the Ring-Expansion Polymerization of Macrocycles with a Bis(Hindered Amino)Disulfide Linker. *Macromolecules* **2021**, *54* (17), 8154–8163. https://doi.org/10.1021/acs.macromol.1c01067.

(9) Watanabe, N.; Ikari, Y.; Kihara, N.; Takata, T. Bridged Polycatenane. *Macromolecules* **2004**, *37* (18), 6663–6666. https://doi.org/10.1021/ma048782u.

(10) Zhang, K.; Lackey, M. A.; Cui, J.; Tew, G. N. Gels Based on Cyclic Polymers. *J. Am. Chem. Soc.* **2011**, *133* (11), 4140–4148. https://doi.org/10.1021/ja111391z.

(11) Mena-Hernando, S.; Pérez, E. M. Mechanically Interlocked Materials. Rotaxanes and Catenanes beyond the Small Molecule. *Chem. Soc. Rev.* **2019**, *48* (19), 5016–5032. https://doi.org/10.1039/C8CS00888D.

(12) Hu, L.; Lu, C.-H.; Willner, I. Switchable Catalytic DNA Catenanes. *Nano Lett.* **2015**, *15* (3), 2099–2103. https://doi.org/10.1021/nl504997q.

(13) Duda, R. L.; Teschke, C. M. The Amazing HK97 Fold: Versatile Results of Modest Differences. *Current Opinion in Virology* **2019**, *36*, 9–16. https://doi.org/10.1016/j.coviro.2019.02.001.

(14) Helgstrand, C.; Wikoff, W. R.; Duda, R. L.; Hendrix, R. W.; Johnson, J. E.; Liljas, L. The Refined Structure of a Protein Catenane: The HK97 Bacteriophage Capsid at 3.44 Å Resolution. *Journal of Molecular Biology* **2003**, *334* (5), 885–899. https://doi.org/10.1016/j.jmb.2003.09.035.

(15) Nomidis, S. K.; Carlon, E.; Gruber, S.; Marko, J. F. DNA Tension-Modulated Translocation and Loop Extrusion by SMC Complexes Revealed by Molecular Dynamics Simulations. *Nucleic Acids Research* **2022**, *50* (9), 4974–4987. https://doi.org/10.1093/nar/gkac268.
18

(16) Zhao, H.; Shu, L.; Qin, S.; Lyu, F.; Liu, F.; Lin, E.; Xia, S.; Wang, B.; Wang, M.; Shan, F.; Lin, Y.; Zhang, L.; Gu, Y.; Blobel, G. A.; Huang, K.; Zhang, H. Extensive Mutual Influences of SMC Complexes Shape 3D Genome Folding. *Nature* **2025**. https://doi.org/10.1038/s41586-025-08638-3.

(17) Nuebler, J.; Fudenberg, G.; Imakaev, M.; Abdennur, N.; Mirny, L. A. Chromatin Organization by an Interplay of Loop Extrusion and Compartmental Segregation. *Proc. Natl. Acad. Sci. U.S.A.* **2018**, *115* (29). https://doi.org/10.1073/pnas.1717730115.

(18) Rao, S. S. P.; Huang, S.-C.; Glenn St Hilaire, B.; Engreitz, J. M.; Perez, E. M.; Kieffer-Kwon, K.-R.; Sanborn, A. L.; Johnstone, S. E.; Bascom, G. D.; Bochkov, I. D.; Huang, X.; Shamim, M. S.; Shin, J.; Turner, D.; Ye, Z.; Omer, A. D.; Robinson, J. T.; Schlick, T.; Bernstein, B. E.; Casellas, R.; Lander, E. S.; Aiden, E. L. Cohesin Loss Eliminates All Loop Domains. *Cell* **2017**, *171* (2), 305-320.e24. https://doi.org/10.1016/j.cell.2017.09.026.

(19) Takaki, R.; Savich, Y.; Brugués, J.; Jülicher, F. Active Loop Extrusion Guides DNA-Protein Condensation. *Phys. Rev. Lett.* **2025**, *134* (12), 128401. https://doi.org/10.1103/PhysRevLett.134.128401.

(20) He, P.; Katan, A. J.; Tubiana, L.; Dekker, C.; Michieletto, D. Single-Molecule Structure and Topology of Kinetoplast DNA Networks. *Phys. Rev. X* **2023**, *13* (2), 021010. https://doi.org/10.1103/PhysRevX.13.021010.

(21) Hajduk, S.; Ochsenreiter, T. RNA Editing in Kinetoplastids. *RNA Biology* **2010**, *7* (2), 229–236. https://doi.org/10.4161/rna.7.2.11393.

(22) Rauch, C. A.; Perez-Morga, D.; Cozzarelli, N. R.; Englund, P. T. The Absence of Supercoiling in Kinetoplast DNA Minicircles. *The EMBO Journal* **1993**, *12* (2), 403–411. https://doi.org/10.1002/j.1460-2075.1993.tb05672.x.

(23) Krajina, B. A.; Zhu, A.; Heilshorn, S. C.; Spakowitz, A. J. Active DNA Olympic Hydrogels Driven by Topoisomerase Activity. *Phys. Rev. Lett.* **2018**, *121* (14), 148001. https://doi.org/10.1103/PhysRevLett.121.148001.

(24) Mir, M.; Bickmore, W.; Furlong, E. E. M.; Narlikar, G. Chromatin Topology, Condensates and Gene Regulation: Shifting Paradigms or Just a Phase? *Development* **2019**, *146* (19), dev182766. https://doi.org/10.1242/dev.182766.

(25) Kim, Y.; Joo, S.; Kim, W. K.; Jeon, J.-H. Active Diffusion of Self-Propelled Particles in Flexible Polymer Networks. *Macromolecules* **2022**, *55* (16), 7136–7147. https://doi.org/10.1021/acs.macromol.2c00610.

(26) Sorichetti, V.; Hugouvieux, V.; Kob, W. Dynamics of Nanoparticles in Polydisperse Polymer Networks: From Free Diffusion to Hopping. *Macromolecules* **2021**, *54* (18), 8575–8589. https://doi.org/10.1021/acs.macromol.1c01394.

(27) Xu, Z.; Dai, X.; Bu, X.; Yang, Y.; Zhang, X.; Man, X.; Zhang, X.; Doi, M.; Yan, L.-T. Enhanced Heterogeneous Diffusion of Nanoparticles in Semiflexible Networks. *ACS Nano* **2021**, *15* (3), 4608–4616. https://doi.org/10.1021/acsnano.0c08877.

(28) Shrestha, U. M.; Han, L.; Saito, T.; Schweizer, K. S.; Dadmun, M. D. Mechanism of Soft Nanoparticle Diffusion in Entangled Polymer Melts. *Macromolecules* **2020**, *53* (17), 7580–7589. https://doi.org/10.1021/acs.macromol.0c00870.

(29) Gelb, L. D.; Gubbins, K. E. Pore Size Distributions in Porous Glasses: A Computer Simulation Study. *Langmuir* **1999**, *15* (2), 305–308. https://doi.org/10.1021/la9808418.

(30) Torquato, S.; Avellaneda, M. Diffusion and Reaction in Heterogeneous Media: Pore Size Distribution, Relaxation Times, and Mean Survival Time. *The Journal of Chemical Physics* **1991**, *95* (9), 6477–6489. https://doi.org/10.1063/1.461519.

(31) Sorichetti, V.; Hugouvieux, V.; Kob, W. Determining the Mesh Size of Polymer Solutions via the Pore Size Distribution. *Macromolecules* **2020**, *53* (7), 2568–2581. https://doi.org/10.1021/acs.macromol.9b02166.
(32) Kremer, K.; Grest, G. S. Dynamics of Entangled Linear Polymer Melts: A Molecular-Dynamics Simulation. *The Journal of Chemical Physics* **1990**, *92* (8), 5057–5086. https://doi.org/10.1063/1.458541.
(33) Weeks, J. D.; Chandler, D.; Andersen, H. C. Role of Repulsive Forces in Determining the Equilibrium Structure of Simple Liquids. *The Journal of Chemical Physics* **1971**, *54* (12), 5237–5247. https://doi.org/10.1063/1.1674820.
(34) Zhu, Y.-L.; Liu, H.; Li, Z.-W.; Qian, H.-J.; Milano, G.; Lu, Z.-Y. GALAMOST: GPU-Accelerated Large-Scale Molecular Simulation Toolkit. *J. Comput. Chem.* **2013**, *34* (25), 2197–2211. https://doi.org/10.1002/jcc.23365.
(35) Calladine, C. R. Buckminster Fuller's "Tensegrity" Structures and Clerk Maxwell's Rules for the Construction of Stiff Frames. *International Journal of Solids and Structures* **1978**, *14* (2), 161–172. https://doi.org/10.1016/0020-7683(78)90052-5.
(36) Thorpe, M. F. CONTINUOUS DEFORMATIONS IN RANDOM NETWORKS.
(37) Phillips, J. C. TOPOLOGY OF COVALENT NON-CRYSTALLINE SOLIDS II: MEDIUM-RANGE ORDER IN CHALCOGENIDE ALLOYS AND A-Si(Ge).
(38) Micoulaut, M. Relaxation and Physical Aging in Network Glasses: A Review. *Rep. Prog. Phys.* **2016**.
(39) Kane, C. L.; Lubensky, T. C. Topological Boundary Modes in Isostatic Lattices. *Nature Phys* **2014**, *10* (1), 39–45. https://doi.org/10.1038/nphys2835.
(40) Mao, X.; Lubensky, T. C. Maxwell Lattices and Topological Mechanics. *Annu. Rev. Condens. Matter Phys.* **2018**, *9* (1), 413–433. https://doi.org/10.1146/annurev-conmatphys-033117-054235.
(41) Zhou, D.; Zhang, L.; Mao, X. Topological Edge Floppy Modes in Disordered Fiber Networks. *PHYSICAL REVIEW LETTERS* **2018**.
(42) Zhou, D.; Zhang, L.; Mao, X. Topological Boundary Floppy Modes in Quasicrystals. **2019**.